\documentclass[preprint,preprintnumbers,amsmath,amssymb]{revtex4}

\usepackage[pdftex]{graphicx}
 
\usepackage{dcolumn}% Align table columns on decimal point
\usepackage{bm}% bold math
 
\begin{document}

\title{Formation and Rupture of Ca$^{2+}$ Induced Pectin Biopolymer Gels}
\vspace{0.6cm}

\author{Rajib Basak}
 \email{rajib@rri.res.in}
\affiliation{Raman Research Institute, C. V. Raman Avenue, Sadashivanagar, Bangalore 560080, India}
\author{Ranjini Bandyopadhyay}
\email{ranjini@rri.res.in}
\affiliation{Raman Research Institute, C. V. Raman Avenue, Sadashivanagar, Bangalore 560080, India}

\vspace{0.6cm}
%Please do not change this text.
\date{\today}
\begin{abstract}
When calcium salts are added to an aqueous solution of polysaccharide pectin, ionic cross-links form between pectin chains, giving rise to a gel network in dilute solution. In this work, dynamic light scattering (DLS) is employed to study the microscopic dynamics of the fractal aggregates (flocs) that constitute the gels, while rheological measurements are performed to study the process of gel rupture. As calcium salt concentration is increased, DLS experiments reveal that the polydispersities of the flocs increase simultaneously with the characteristic relaxation times of the gel network. Above  a critical salt concentration, the flocs become interlinked to form a reaction-limited fractal gel network. Rheological studies demonstrate that the limits of the linear rheological response and the critical stresses required to rupture these networks both decrease with increase in salt concentration. These features indicate that the ion-mediated pectin gels studied here lie in a `strong link' regime that is characterised by inter-floc links that are stronger than intra-floc links. A  scaling analysis of the experimental data presented here demonstrates that the elasticities of the individual fractal flocs exhibit power-law dependences on the added salt concentration. We conclude that when pectin and salt concentrations are both increased, the number of fractal flocs of pectin increases simultaneously with the density of crosslinks, giving rise to very large values of the bulk elastic modulus
\end{abstract}

\maketitle

\section{Introduction}

%Please use \dag to cite the ESI in the main text of the article.
%If you article does not have ESI please remove the the \dag symbol from the title and the above footnotetext.

Pectins, extracted from plant cell walls, are complex polysaccharides. Pectin is widely used in the food industry as a gelling agent, thickener, emulsifier and stabilizer \cite{handa_review}. Pectin consists primarily of linear chains of partly methyl esterified $\alpha$-d-galacturonic acid, interrupted and bent in places by rhamnose units. Depending upon the degree of esterification of the carboxyl group of galacturonic acid with methanol, pectin can be categorized into two types. The species of pectin with degrees of methylation (molar ratio of methanol to galacturonic acid) higher than 50\% is called high methoxyl pectin, while pectin samples with degrees of methylation less than 50\% are called low methoxyl pectin. The gelation of low methoxyl (LM) pectin depends upon the pH of the system and the presence of Ca$^{2+}$ or other multivalent ions. At low pH and in the absence of Ca$^{2+}$ ions, LM pectin can form gels \textit{via} non-ionic associations based on hydrophobic interactions and the formation of hydrogen bonds \cite{gilsenan_carbohypoly}. In the presence of divalent cations like Ca$^{2+}$, pectin gels can form \textit{via} Ca$^{2+}$ ion bridges between two carboxyl groups of two different chains. This mechanism is popularly known as egg-box mechanism and was first observed in Ca$^{2+}$ induced alginate biopolymer gels \cite{braccini_biomacro,ravanat_biopoly,morris_carbohy}. Recently, small-angle x-ray scattering (SAXS), dynamic light scattering (DLS) and diffusive wave spectroscopy (DWS) were employed to report that rod-like junctions, both single ion junctions and those comprising of multi-unit egg-box structures, form in Ca$^{2+}$ induced biopolymer gels \cite{ventura_carbohypoly,larobina_softmatter}. Time resolved light scattering measurements were employed to show that these hierarchical crosslinking structures result in two relaxation modes in light scattering measurements, associated with discontinuous rearrangement events that restructure the gel network under mechanical stress \cite{larobina_softmatter}. The kinetics of formation and aging of alginate gels with controlled permeation of CaCl$_{2}$ was studied by a photon correlation imaging technique \cite{piazza_softmatter}.  This work reported a non-diffusive behaviour of the gelling front and demonstrated that the aging behaviour of these gels is strongly reminiscent of colloidal gels. \\
   \indent Pectins are major components of the primary plant cell wall, where they sometimes comprise 30-35\% of the cell wall dry weight \cite{pell_plant}. Ca$^{2+}$ ions in the plant cell wall form pectin gel structures {\it via} ionic crosslinking. Ca$^{2+}$, which  plays a crucial role in determining the structural rigidity of the cell wall, is a crucial regulator of growth and development in plants \cite{jones_botany}. Low concentrations of Ca$^{2+}$ result in cell walls that are more flexible and therfore easily ruptured, while high concentrations of Ca$^{2+}$ increase the rigidity the wall, while making it less plastic \cite{hepler_plant}. To understand cell wall behavior, it is important to understand the formation and rupture of pectin gels. Furthermore, knowledge of the mechanical properties of pectin gels has practical implications in the food industry, mainly in the production of jams, preservtaives {\it etc.} \cite{may_carbo}. Ion-induced hydrogels are also used for many application related to pharmaceuticals and for tissue engineering, glucose transport, cell encapsulation, drug delivery {\it etc} \cite{sriam, surita_cis, surita_aps,surita_bio}.  The various applications of pectin gels, therefore, necessitate a detailed study of their formation and mechanical stability under different conditions.  \\
\indent The mechanical properties of ion induced hydrogels can be improved by the addition of block copolymers in bioengineering applications \cite{surita_biomacro}. Calcium-ion induced gelation of LM pectin therefore depends upon various parameters like pectin concentration, salt concentration, the degree of methylation of pectin molecules, the pH and temperature of the solution, the presence of additives, {\it etc}. The effects of these parameters on the gelation process of pectin solutions have been studied by rheological methods \cite{garnier_carbohy,yoo_foodhydro,durand_ijb_macro,clark_advpoly,axelos_prl,looten_foodhydro,cordoso_foodhydro,fraeye_foodhydro}, DLS \cite{narayanan_jcis,kjoniksen_europoly}, nuclear magnetic resonance (NMR) \cite{dobies_actaphysica} and circular dichroism \cite{morris_molbio,thom_carbores}. Conductometric and potentiometric studies show that decreasing the degree of methylation of pectin molecules and the ionic strength of the solution can increase the affinity of pectin molecules towards Ca$^{2+}$ ions. This results in a decrease in the amount of calcium chloride required to obtain a sol-gel transition \cite{garnier_carbohy,gar_carbohy}. The sol-gel transition in solutions of pectin and other biopolymer molecules can also be triggered by irradiating the samples with ultraviolet radiation \cite{raghavan_langmuir}. Durand \textit{et. al.} \cite{durand_ijb_macro} reported the gelation times of pectin solutions for various pectin concentrations and stoichiometric ratios of salt and pectin and estimated sol-gel diagrams for calcium-pectin systems for a range of calcium levels, solution temperatures and pectin concentrations. Lootens \textit{et. al.} \cite{looten_foodhydro} showed that the lowering of the solution pH weakens the Ca$^{2+}$ induced gel formation process. DLS measurements exhibit an increase in the characteristic relaxation times of pectin solutions when the concentration of Ca$^{2+}$ in solution is increased \cite{narayanan_jcis}.    \\
		\indent In this work, the microscopic dynamics of gelling LM pectin solutions are investigated by measuring intensity autocorrelation functions in DLS measurements. The relaxation timescales of the samples, which are extracted from the decays of the autocorrelation functions obtained from the samples before gelation, are compared with their bulk viscosities. Next, the effects of applied stresses on pectin gels of various strengths are studied and the critical stresses required to break the gel networks are estimated using rheological measurements. The dependence of the critical stress on the pectin gel strength is investigated systematically by changing the CaCl$_{2}$ concentration. The results are explained in terms of the fractal nature of the pectin gel structure using a scaling theory that was first introduced by Shih \textit{et. al.} to describe the elastic properties of colloidal gels \cite{shih_prl}. It should be noted that while the colloidal gels investigated in this work \cite{shih_prl} were formed by increasing the volume fraction of boehmite alumina particles of Catapal and Dispal respectively, our work studies the process of ion-mediated gelation of pectin solutions that is triggered by increasing the CaCl$_{2}$ concentration in solution. Analysis of our experimental data shows a strong power-law dependence of the elasticity of individual pectin flocs on the added salt concentration. Our data suggests that when pectin and salt concentrations are both increased, the number of aggregates increases simultaneously with the density of crosslinks and that the inter-aggregate links are stronger than the intra-aggregate links.

%---- FOLLOWING SECTION IS THE PART OF THE TEMPLATE BY RSC
%\subsection{This is the subsection heading style}
%Section headings can be typeset with and without numbers.\cite{Abernethy2003}

%\subsubsection{This is the subsubsection style.~~} These headings should end in a full point.  

%\paragraph{This is the next level heading.~~} For this level please use \texttt{\textbackslash paragraph}. These headings should also end in a full point.

%------------------------------------------------------------------
\section{ MATERIALS AND METHODS}
\indent \textbf{2.1. Sample Preparation.} Pectin, a 20-34\% esterified potassium salt, extracted from citrus fruit and calcium chloride anhydrous (molecular weight 111 g/mol) are purchased from Sigma-Aldrich and used as received without further purification. Stock solutions of calcium chloride of appropriate concentrations are prepared. Pure pectin solutions are prepared by dissolving appropriate amounts of pectin in deionized and distilled Millipore water and by stirring the mixture vigorously with a magnetic stirrer. Stock solutions of CaCl$_{2}$ of known concentrations are mixed with pure pectin solutions to prepare gel samples of several strengths. The mixtures are stirred overnight before loading the samples for the experiments. All experiments are performed at 25$^{\circ}$C.\\ 

 \indent \textbf{2.2. Dynamic light Scattering.} DLS measurements are performed with the help of BIC (Brookhaven Instruments Corporation) BI-200SM spectrometer using a 150 mW solid state laser (NdYVO$_{4}$, Spectra Physics Excelsior) with an emission wavelength of 532 nm as the light source. A temperature controller (Polyscience Digital) is used to control the temperature of the sample cell.  The intensity auto-correlation function  \cite{berne_dls}, $G^{(2)}(\tau) = \frac{<I(0)I(\tau)>}{<I(0)>^{2}} =  1+ A|g^{(1)}(\tau)|^{2}$,
where $I(\tau)$ is the intensity at time $\tau$, $g^{(1)}(\tau)$ is the normalized electric field auto-correlation function, \textit{A} is the coherence  factor and the angular bracket $< >$ represents an average over time, is measured with a Brookhaven BI-9000AT Digital autocorrelator. If the sample is monodisperse and dilute,  $g^{(1)}(\tau) \sim \exp(-\Gamma\tau)$, where $\Gamma$ is the relaxation rate \cite{berne_dls}.  If the scatterers are polydisperse in size, $C(\tau) = G^{(2)}(\tau)\sim \exp(-\Gamma\tau)^{\beta}$, with  $\beta$ $\leq 1$. A lower estimated value of $\beta$ indicates a higher scatterer polydispersity. In the experiments reported here, $C(\tau)$ is measured at different scattering angles $\theta$ corresponding to different scattering wave vectors $q={\frac{4\pi{n}}{\lambda}}\sin (\theta/2)$, where $\lambda$ is the wavelength  of light and $n$ is the refractive index of the solvent.

\indent \textbf{2.3. Static light Scattering.} Static light scattering (SLS) experiments are also performed in the BI-200SM spectrometer. In these experiments, the total intensity $I(q)$ is measured at different scattering wave vectors $q$ that lie between 0.001 $nm^{-1}$ and 0.01 $nm^{-1}$. 

\indent \textbf{2.4. Rheology.}
All the rheological measurements are performed in a commercial modular compact rheometer Anton Paar MCR 501 after loading a fresh sample for every run in a concentric cylinder geometry CC17. This geometry has a gap of 0.71 mm, an effective length of 24.99 mm and requires a sample volume of 4.72 ml for every run. The sample temperature is controlled by a water circulation unit Viscotherm VT2. The rheological data is acquired using the Rheoplus software (version 3.40) provided by the manufacturer. The time evolution experiments are performed by loading the samples in the rheometer immediately after mixing the CaCl$_{2}$ solution with the pure pectin solution. In the creep test, the time-dependence of the strain that develops due to the application of a constant stress on the sample is recorded, while in the recovery test, the time-dependence of the strain is measured after the removal of the applied stress. In the oscillatory amplitude sweep test, an oscillatory strain with increasing amplitude $\gamma$ and constant angular frequency $\omega$ is applied and the storage modulus ($G^{\prime}$) and loss modulus ($G^{\prime\prime}$) are measured. 

\indent \textbf{2.5. Cryo-SEM (cryogenic scanning electron microscopy) measurements.} For the cryo-SEM experiments, the samples are first cryo-fractured with the help of liquid nitrogen. The samples are then transferred to a PP3000T cryo unit (Quorum Technologies) and cut with a cold knife. The fractured and cut samples are sublimated at $-130^{\circ}$C for 5 minutes and then sputtered with platinum inside the unit. Imaging of the samples is performed using a Zeiss Ultra Plus cryo-SEM setup at a temperature of $-160^{\circ}$C.

\section{Results and Discussions}
\subsection{Gel Formation:}
\begin{figure}[!t]
\begin{center}
\includegraphics[width=3.5in]{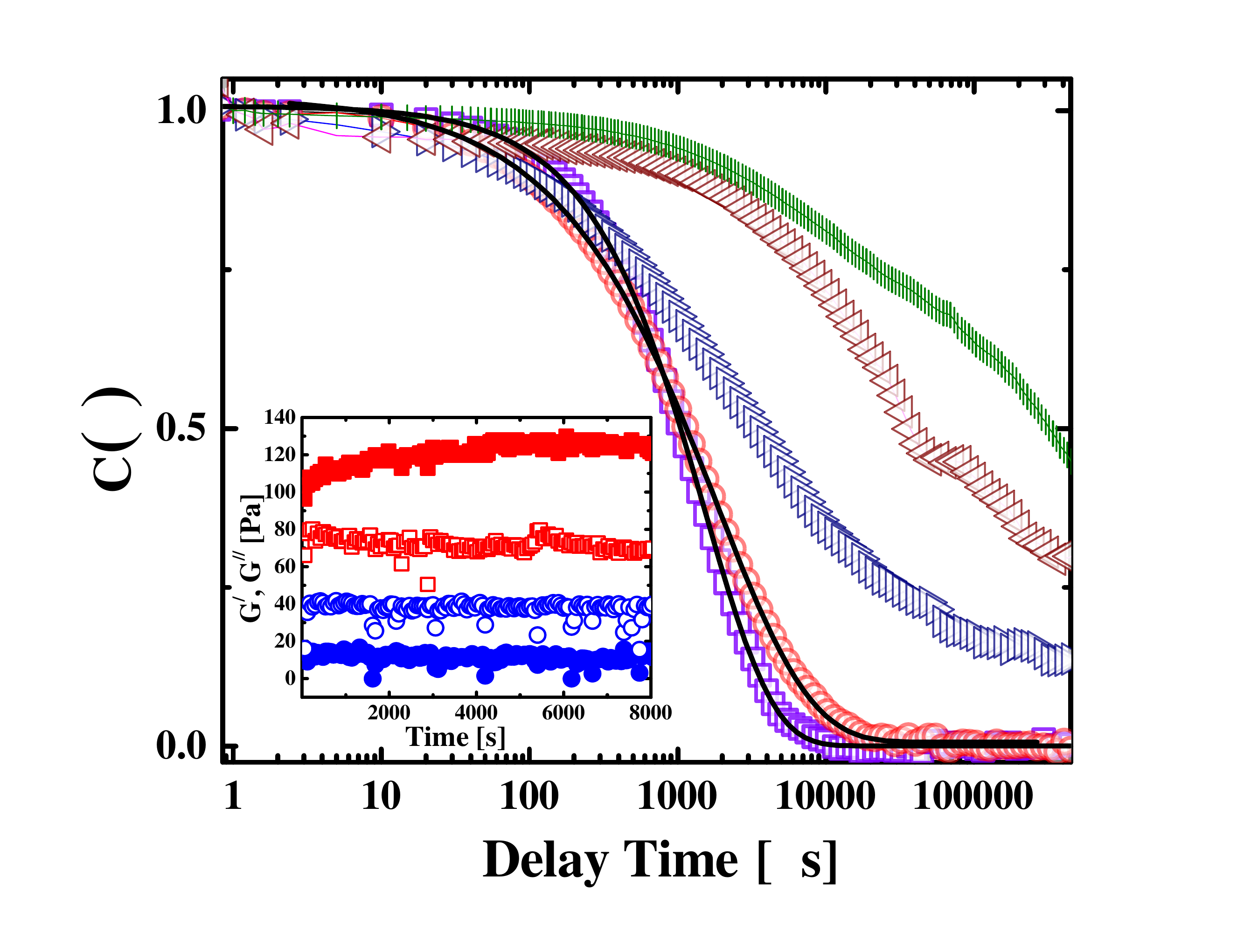}
\caption{Plots of the normalized autocorrelation functions $C(\tau)$ \textit{vs.} delay time $\tau$ at scattering angle $\theta$ = $90^{\circ}$ for 2.5 g/L pectin solution at 25$^{\circ}$C for different CaCl$_{2}$ concentrations: 0 mM ($\square$), 2 mM ($\circ$), 6 mM ($\triangleright$), 8 mM ($\triangleleft$), 10 mM ($\mid$). The stretched exponential fits to the data for 0 mM and 2 mM CaCl$_{2}$ concentrations are shown by solid black lines. In the inset, the time evolutions of the storage moduli $G^{\prime}$ (solid symbols) and the loss moduli $G^{\prime\prime}$ (hollow symbols), after the addition of 2 mM CaCl$_{2}$ ($\circ$) and 6 mM CaCl$_{2}$ ($\square$) to 2.5 g/L pectin solution at 25$^{\circ}$C, are plotted. The experiments are performed in oscillatory measurements at an applied strain 0.5\% and at an angular frequency 1 rad/s.}
\label{FIG 1}
\end{center}
\end{figure}
 Systematic DLS studies are performed with aqueous solutions of pectin of concentration 2.5 g/L   with different concentrations of added CaCl$_{2}$ salt. Fig. 1 shows the plots of the normalized intensity-intensity autocorrelation functions $C(\tau)$ \textit{vs.} delay times $\tau$ for different CaCl$_{2}$ concentrations. It is observed that for pectin solutions with low or no salt, the $C(\tau)$ plots show complete decays. To estimate the distributions of relaxation rates, the correlation decays are fitted to stretched exponential functions 

	\[C(\tau) \sim \exp(-\Gamma_{R}\tau)^{\beta} \tag {1}
\]

\noindent  where $\Gamma_{R}$, the relaxation rate, is the inverse of a relaxation time $\tau_{R}$ and $\beta$ $\leq 1$ is the stretching exponent. The mean relaxation time $<\tau_{R}>$ for a particular scattering angle is estimated using the following relation 
	 \[<\tau_{R}> = \tau_{R}/\beta*\Gamma({1/\beta}) \tag {2}
 \]
\noindent where ${\Gamma}$  is the Euler Gamma function \cite{lindsay_jcp}.\\
\begin{figure}
\begin{center}
\includegraphics[width=3.5in]{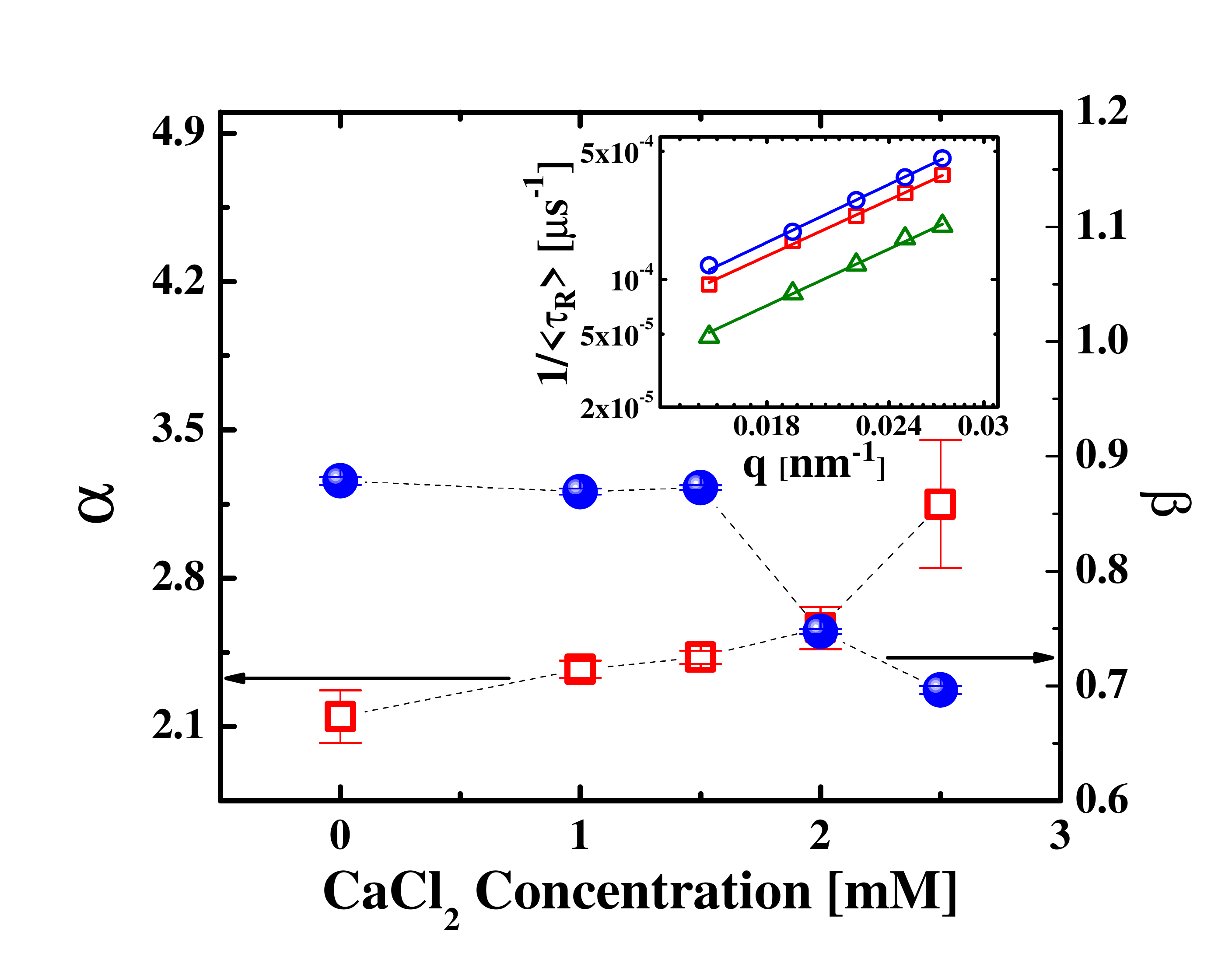}
\caption{ Plots of the power law exponents $\alpha$ (extracted from the fits of the mean relaxation time $<\tau_{R}>$ to Eqn. 3 and represented by $\square$) (Eqn. 3) and the stretched exponents $\beta$ ($\bullet$) (extracted from fits of $C(\tau)$ \textit{vs.} $\tau$  to Eqn. 1 and represented by solid circles),  \textit{vs.} CaCl$_{2}$ concentration at 25$^{\circ}$C for 2.5 g/L pectin solutions. The inset shows the plots of $1/<\tau_{R}>$ $vs.$ $q$ with different CaCl$_{2}$ concentrations : 0 mM ($\circ$), 1 mM ($\square$), 2 mM ($\triangle$) and the corresponding fits to Eqn. 3 (fits are represented by solid lines).}
\label{FIG 2}
\end{center}
\end{figure}
\indent	 As the CaCl$_{2}$ concentration is increased, a systematic slowdown in the decay times of the correlation functions is observed, with the correlation plots eventually showing incomplete decays at the highest salt concentrations. The observed transition of the sample to a regime of non-ergodic dynamics beyond a threshold salt concentration marks the onset of the gelation process \cite{megen_pra}. The concentration of CaCl$_{2}$, above which an incomplete decay in $C(\tau)$ is observed in the present experiments, is assigned as the critical concentration `$C_{cr}$' of CaCl$_{2}$ salts required for the gelation of pectin solutions. By studying the measured autocorrelation decays, $C_{cr}$ is estimated to lie between 3.5-4 mM for the 2.5 g/L pectin solutions. Gel formation, due to an increase in CaCl$_{2}$ concentration above $C_{cr}$, is confirmed by plotting the time evolution of the rheological moduli. It is seen from the inset of Fig. 1 that when the CaCl$_{2}$ concentration (2 mM data represented by circles) lies below $C_{cr}$,  the sample shows predominantly viscous response to small oscillatory strains at the angular frequency that has been probed here (1 rad/sec), with the loss modulus $G^{\prime\prime}$ always higher than the storage modulus $G^{\prime}$. When the CaCl$_{2}$ concentration is above $C_{cr}$ (6 mM data represented by squares), $G^{\prime} >> G^{\prime\prime}$ from the start of the measurement. The structure of the sample with 6 mM salt is, therefore, far more rigid than the one with 2 mM salt, with gelation being initiated immediately after the addition of salt.\\

\begin{figure}
\begin{center}
\includegraphics[width=3.5in]{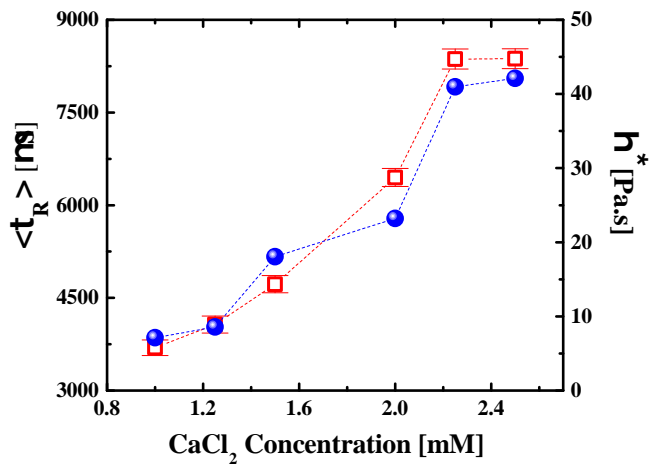}
\caption{ Plot of the mean relaxation time $<\tau_{R}>$ ($\square$) and complex viscosity $\eta ^{*}$ ($\bullet$) \textsl{vs.} CaCl$_{2}$ concentration for gelling pectin solutions (concentration of pectin is kept fixed at 2.5 g/L) at 25$^{\circ}$C.}
\label{FIG 3}
\end{center}
\end{figure}

 \indent Changes in the wave vector dependence of the sample dynamics due to the addition of CaCl$_{2}$ are explored by studying the fits to the stretched exponential correlation decays (Eqn. 1) at different scattering angles. $C(\tau)$ {\it vs.} $\tau$ data for several scattering wavevectors is shown in Fig. S1 of Supporting Information. In the inset of Fig. 2, the inverse of the mean relaxation times $<\tau_{R}>$, extracted from the stretched exponential fits of $C(\tau)$ for samples with CaCl$_{2}$ concentrations below $C_{cr}$, are plotted \textit{vs.} scattering wave vector $q$. This data is fit to the relation 
 
	 \[1/<\tau_{R}>~\approx~q^{\alpha} \tag{3}
 \]
\noindent  The power law exponents $\alpha$ (squares), extracted from  fits to Eqn. 3, and the stretching exponents $\beta$ (circles), obtained from fits to Eqn. 1, are plotted in Fig. 2 for different CaCl$_{2}$ concentrations. At low CaCl$_{2}$ concentrations, $\alpha$ $\approx$ 2 and $\beta$ $\approx$ 0.9. This indicates the presence of pectin flocs of approximately equal sizes diffusing freely in the solution. With increase in CaCl$_{2}$ concentration, $\alpha$ shows a gradual increase from 2. This points to an increase in the spatial inhomogeneity of the system. It is observed that the polydispersity in the floc sizes simultaneously increases with CaCl$_{2}$ concentration, with $\beta$ dropping to values much lower than 1. A stronger-than-diffusive $q$-dependence of the estimated relaxation time and $\beta < 1$  with increasing pectin and divalent salt concentrations have been observed in LM-pectin solutions in earlier reports  \cite{narayanan_jcis,kjoniksen_europoly}. Furthermore,it has been verified by us that the relation $\alpha = \frac{2}{\beta}$ seen in these earlier reports holds in the concentration range 0-2.5 mM investigated here. \\
 
The mean relaxation times $<\tau_{R}>$ of pectin gels (represented by squares) show a monotonic increase when plotted {\it vs.} CaCl$_{2}$ concentration (Fig. 3). The pectin floc sizes therefore increase with increasing salt, causing the system to become increasingly heterogeneous and resulting in the observed dynamical slowdown. Clearly, as the CaCl$_{2}$ concentration is increased, the increased availability of Ca$^{2+}$ ions in pectin solutions  increases the cross-link density between pectin molecules, resulting in the formation of larger and more polydisperse pectin flocs. A further increase in the concentration of CaCl$_{2}$ towards the critical value $C_{cr}$ results in the formation of  links between the pectin chains. Eventually, a percolating gel network is formed at CaCl$_{2}$ concentrations greater than $C_{cr}$. Due to an increase in the number of links with increasing CaCl$_{2}$ concentration, this dynamical slowing down is accompanied by an increase of the complex viscosity $\eta ^{*}$. This is shown in Fig. 3 (circles). The complex viscosity $\eta^{\star}$ is defined as $\eta^{\star} = \frac{{(G^{{\prime}2} + G^{{\prime\prime}2})}^{1/2}}{\omega}$, where G$^{\prime}$ and G$^{\prime\prime}$ are the elastic and viscous moduli respectively, measured at an angular frequency $\omega$ = 1 rad/s and at strain amplitudes $\gamma =$ 0.5\%\cite{macosko_rheology}.\\
\begin{figure}
\begin{center}
\includegraphics[width=3.5in]{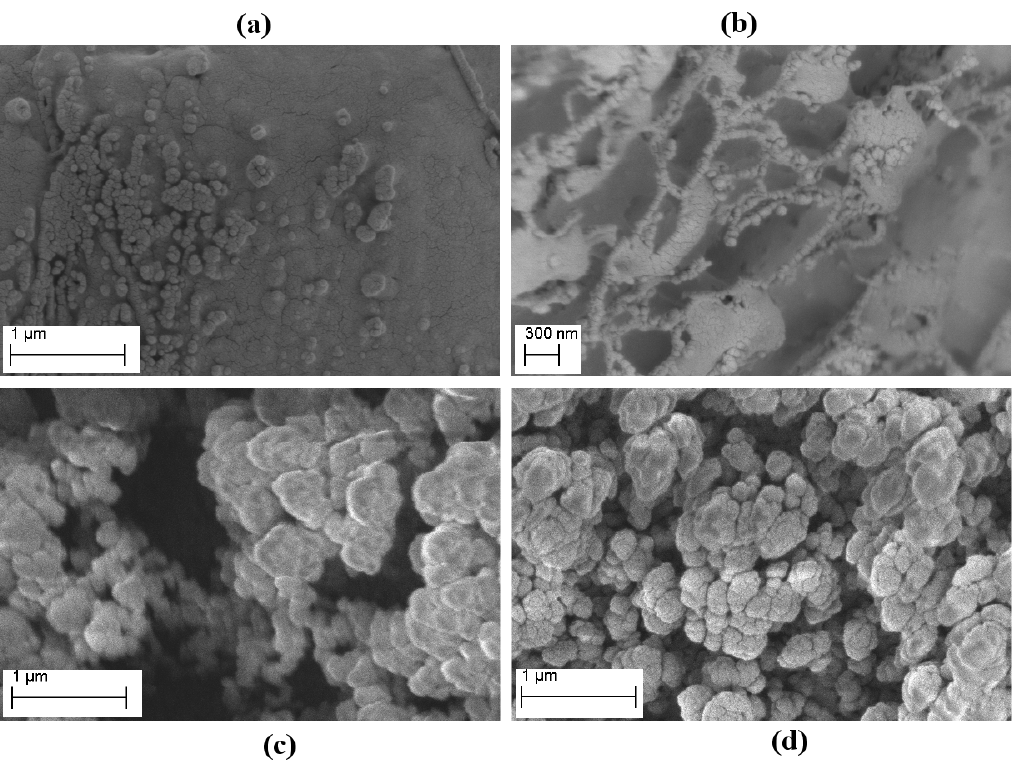}
\caption{ Cryo-SEM image for 2.5 g/L pectin with different CaCl$_{2}$ concentrations: (a) 1 mM, (b) 3.5 mM, (c) 6 mM and (d) 10 mM.}
\label{FIG 4}
\end{center}
\end{figure}
\indent The existence of flocs in pectin solutions and the formation of links between flocculated structures at higher salt concentrations are confirmed from cryo-SEM images  (Fig. 4). In Fig. 4(a), which shows  an image of a pectin sample with CaCl$_{2}$ concentration below $C_{cr}$, small and polydisperse pectin flocs are observed. Near the critical concentration $C_{cr}$ (3.5 mM CaCl$_{2}$ is added to 2.5 g/L pectin solutions in Fig. 4(b)), an increase in the floc sizes and a development of inter-floc links is observed. The formation of links (CaCl$_{2}$ bridges) between pectin molecules results in gelation and the observed divergence of relaxation time scales seen in Fig. 1. Cryo-SEM images of pectin samples containing CaCl$_{2}$ at even higher concentrations (data for pectin gels with 6 mM and 10 mM CaCl$_{2}$ are plotted in Fig. 4(c) and 4(d) respectively) show the presence of even bigger flocs. It is to be noted here that while the sizes of the pectin flocs do not increase substantially in the salt concentration regime $C >> C_{cr}$, the density of interlinks increases sharply with increasing CaCl$_{2}$ concentration.

The SEM images presented here, which provide direct evidence of the growth and percolation of pectin flocs to form gel networks at $C >> C_{cr}$, are in agreement with the DLS data presented earlier. As pectin gels are formed due to the development of interlinks between flocs of different sizes, static light scattering (SLS) measurements are next performed to confirm the fractal nature of the pectin gels. In these measurements, the scattered intensity $I(q)$ is plotted \textit{vs.} wave vector $q$ for pectin solutions containing CaCl$_{2}$ at concentrations that lie above the critical gelation threshold $C_{cr}$. The fractal structure of the pectin gel results in  power law decays of $I(q$) {\it vs.} $q$ according to the following relation \cite{lin_prsm}
	\[I(q)~\approx~q^{-D} \tag{4}
\]
\begin{figure}
\begin{center}
\includegraphics[width=3.5in]{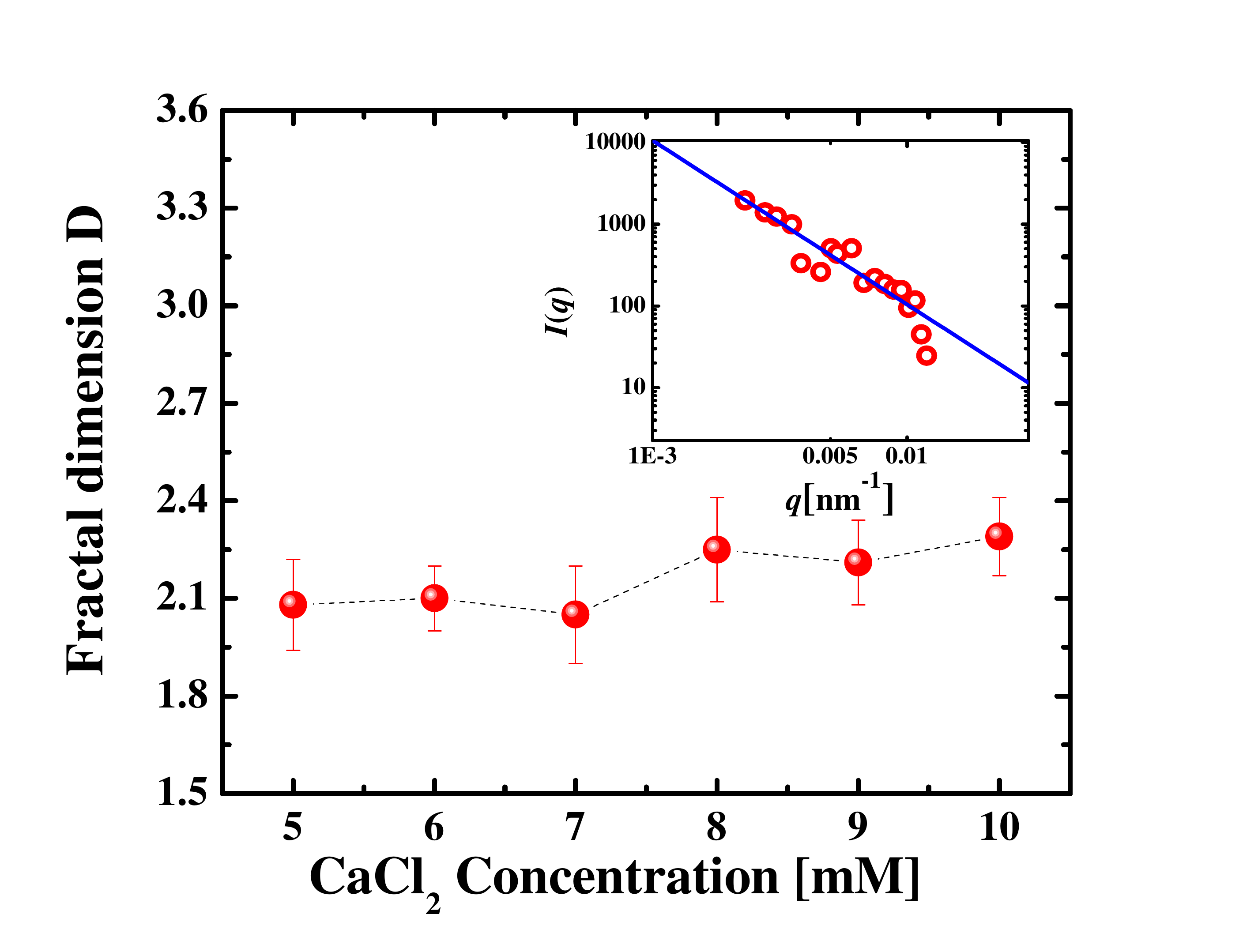}
\caption{ Fractal dimensions D of pectin gels are plotted \textit{vs.} CaCl$_{2}$ concentration for 2.5 g/L pectin solutions containing CaCl$_{2}$ at concentrations greater than $C_{cr}$. In the inset, $I(q)$ \textit{vs.} \textit{q} plot and the corresponding fit to Eqn. 4 (solid line) for a 2.5 g/L pectin solution containing 6 mM CaCl$_{2}$ at 25$^{\circ}$C are plotted.}
\label{FIG 5}
\end{center}
\end{figure}

\noindent where $D$ is the fractal dimension of the gel. $I(q$) \textit{vs.} $q$ data has been recorded for 2.5 g/L pectin solutions containing  CaCl$_{2}$ in the concentration range 5-10 mM. The data for the pectin gel containing 6 mM CaCl$_{2}$ is plotted in the inset of Fig. 5. The exponent of the power-law fit to the decay (solid line in the inset) gives the fractal dimension $D$ of the gel.  The fractal dimensions obtained from power-law decay fits for the entire salt concentration range above $C_{cr}$ investigated here are plotted in Fig. 5. $D$ is always seen to lie between 1.9 and 2.4, suggesting that size-polydisperse pectin gel flocs are formed by a process of three-dimensional reaction-limited aggregation (RLA) in this CaCl$_{2}$ concentration range \cite{ball_prl}. Unlike in globular protein gels, where the aggregation process changes from reaction-limited to diffusion-limited with increase in ionic concentration  \cite{ikeda_lang}, the approximately constant values of $D$ reported here indicates that the interaction mechanism triggering the gelation of pectin does not change within the  experimental range of CaCl$_{2}$ concentrations investigated here. This is in sharp contrast to the ion-mediated gelation of globular proteins, where the probability of collisions between protein molecules and the subsequent formation of interlinks increases  rapidly at higher ionic strengths due to the increased shielding of surface charges on the protein molecules.

\begin{figure}
\begin{center}
\includegraphics[width=3.5in]{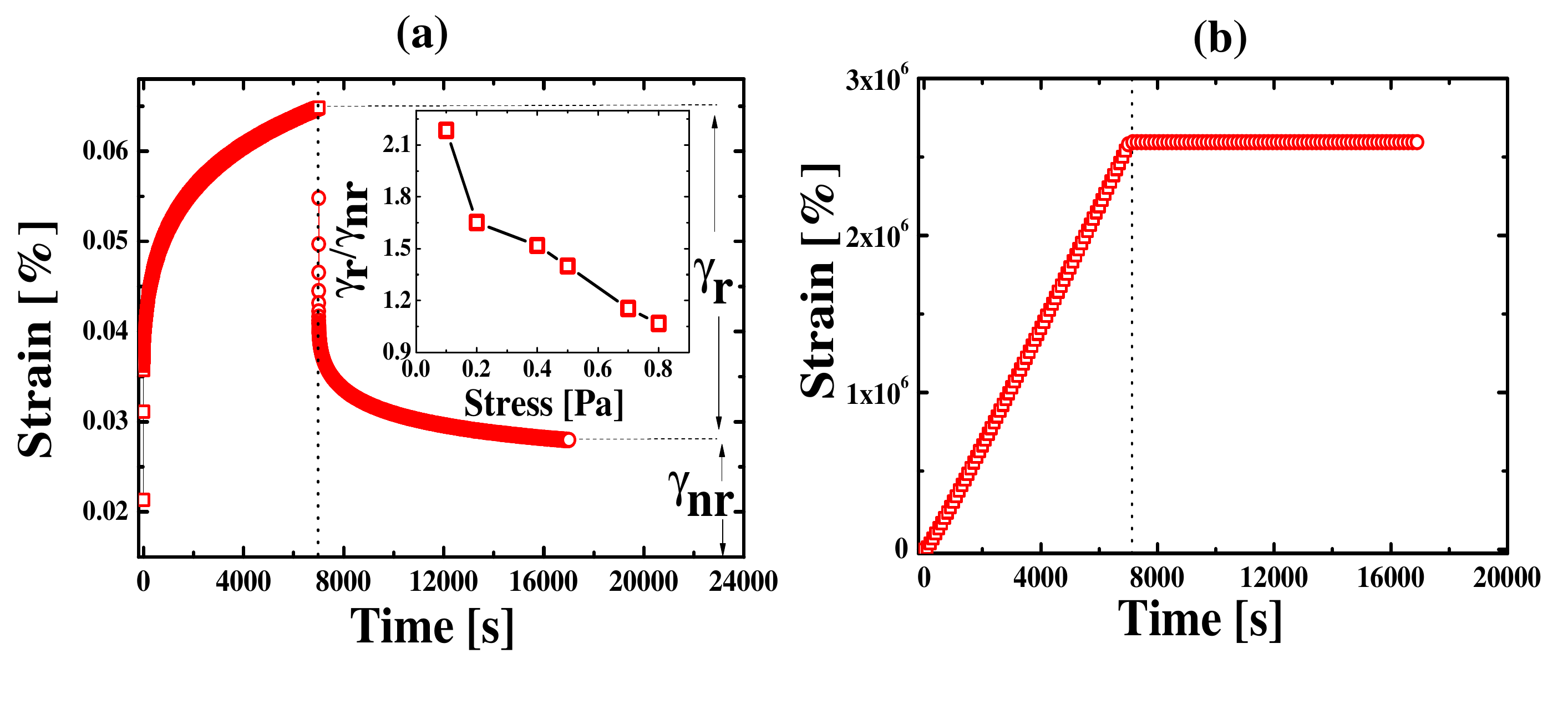}
\caption{ Strain variations during creep and recover tests for 2.5 g/L pectin gel containing 6 mM CaCl$_{2}$ at 25$^{\circ}$C. Data for applied stresses 0.4 Pa and 2 Pa are plotted in (a) and (b) espectively. In the inset of (a), the plot of $\gamma_{r}$/$\gamma_{nr}$  \textsl{vs.} applied stress for 2.5 g/L pectin gel containing 6 mM CaCl$_{2}$ is shown. The dotted line denotes the time at which the applied stress is set to zero.}
\label{FIG 6}
\end{center}
\end{figure}
\subsection{Gel Rupture:}
 To complement the data on pectin gel formation due to the addition of CaCl$_{2}$, rheological experiments are performed to study the rupture of these gels following the application of shear stresses.  Creep and recovery tests are performed to investigate the response of the cross-linked pectin flocs to externally-imposed shear stresses. It is observed that the rheological response of pectin gels is very strongly dependent upon the amplitude of the applied stress.  
 In Fig. 6(a), the development of strain for a pectin gel network, when a low constant stress ($\sigma$ =0.4 Pa) is applied for 7200 seconds to a 2.5 g/L pectin gel with 6 mM CaCl$_{2}$, is plotted. It is seen in Fig. 6(a) that the sample deforms almost immediately after the application of the stress. After an initial instantaneous deformation, the strain continues to develop slowly with time and reaches a value of $\gamma_{M}$ after 7200 seconds. When the applied stress is removed from the sample, it is observed that most of the strain recovers almost immediately. This is followed by a slow and continuous process of strain recovery, with the strain eventually reaching a plateau value $\gamma_{nr}$. Here, $\gamma_{nr}$ is a measure of the irreversible or permanent strain incorporated in the sample during the creep and recovery experiments. The recoverable strain $\gamma_{r}$ in these experiments is calculated as $\gamma_{r}$=$(\gamma_{M}-\gamma_{nr})$. The ratios of the recoverable and non-recoverable strains $(\gamma_{r}$/$\gamma_{nr}$) estimated in these experiments quantify the responses of the gel networks to applied stresses. Creep and recovery plots for different applied stresses are shown in Fig. S2 of Supporting Information, and the ratio $\gamma_{r}$/$\gamma_{nr}$ is calculated for each experiment. It is observed in the inset of Fig. 6(a) that $\gamma_{r}$/$\gamma_{nr}$ decreases with increase in the applied stress. The gel network therefore weakens with increase in applied stress, with a permanent strain being built into the gel structure. This eventually leads to the rupture of the gel network. When the applied stress is above a critical value, the deformation increases linearly during the time period of stress application. This is shown in Fig. 6(b). When the applied stress is removed, the strain does not recover at all. This signals the rupture of the Ca$^{2+}$ induced pectin gel networks  at these high stresses. The applied stress, above which the creep behavior of the type shown in Fig. 6(a) changes to a fluid-like response as seen in Fig. 6(b), is assigned as a critical stress $(\sigma_{critical})$ required for completely rupturing the gel network.  The existence of a critical stress is a common feature of  jammed soft materials and is related to the strength of the junctions between the clusters \cite{stoke_softmatter}. The value of  $\sigma_{critical}$ for a 2.5 g/L pectin gel with 6 mM CaCl$_{2}$ is estimated to be approximately 0.81 Pa. The relevant plots are shown in Fig. S3 of Supporting Information.

\indent Next, the values of $\sigma_{critical}$ are estimated from creep measurements for 2.5 g/L pectin solutions containing several different CaCl$_{2}$ concentrations. This is plotted in Fig. 7(a). It is observed that $\sigma_{critical}$ decreases monotonically with the added CaCl$_{2}$ concentration. To understand the effects of CaCl$_{2}$ on gel elasticity and the observed decrease in $\sigma_{critical}$, amplitude sweep experiments are performed with 2.5 g/L pectin gels with different concentrations of added CaCl$_{2}$. In these experiments, the amplitude of the applied oscillatory strain $\gamma$ is increased logarithmically from 0.1\% to 100\%, keeping the angular frequency $\omega$ constant at 1 rad/second at room temperature. In Fig. 7(b), the elastic modulus $G^{\prime}$ is plotted \textit{vs.} $\gamma$ for 2.5 g/L pectin solutions containing different amounts of CaCl$_{2}$. For the lowest $\gamma$ values, all the plots show a linear regime of elastic modulus, with $G^{\prime} \approx G^{\prime}_{0}$. This is followed by a strongly non-linear response with increasing $\gamma$, where $G^{\prime}$ decreases rapidly from $G^{\prime}_{0}$. The loss modulus $G^{\prime\prime}$, measured simultaneously with $G^{\prime}$, shows a peak at a certain strain $\gamma_{ys}$ (Fig. S4 of Supporting Information). $\gamma_{ys}$ is a measure of the yield strain, above which  $G^{\prime}$ and $G^{\prime\prime}$ both show approximately power-law decreases.\\
\begin{figure}
\begin{center}
\includegraphics[width=3.5in]{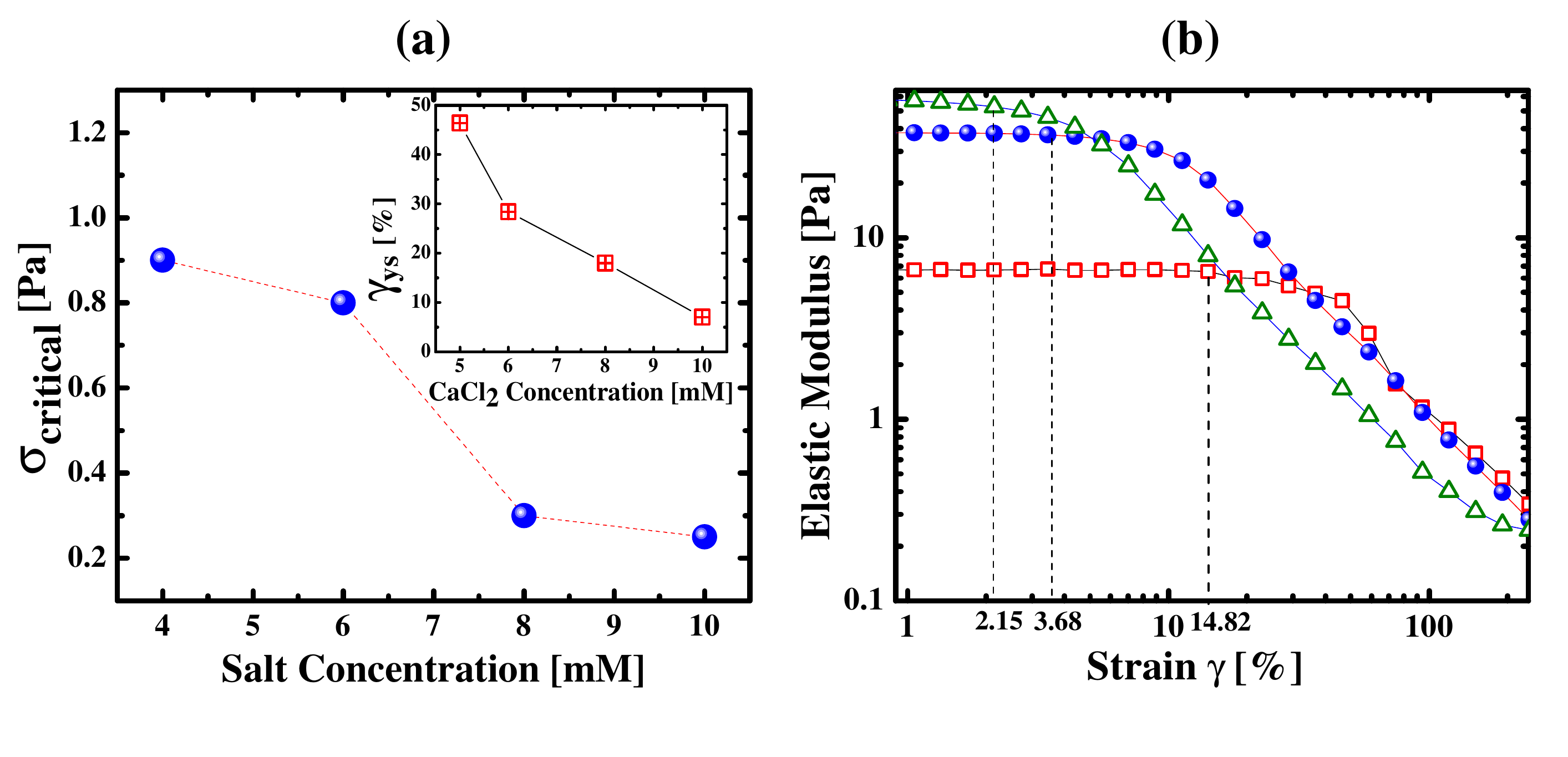}
\caption{ (a) Plot of critical stress $(\sigma_{critical})$ (denoted by $\bullet$) \textit{vs.} CaCl$_{2}$ concentration for 2.5 g/L pectin solutions with added salt. In the inset, a plot of yield strain $\gamma_{ys}$  for 2.5 g/L pectin solutions is shown \textit{vs.} CaCl$_{2}$ concentration. (b) Plot of $G^{\prime}$ \textit{vs.} oscillatory strain amplitude $\gamma$ (angular frequency $\omega$ fixed at 1 rad/s) for 2.5 g/L pectin solutions with CaCl$_{2}$ concentrations of 5mM ($\square$), 8mM ($\bullet$) and 10mM ($\triangle$) is shown. The start of non-linearity is shown by vertical dotted lines.}
\label{FIG 7}
\end{center} 
\end{figure}
\indent The limit of linear rheological response (or the limit of linearity), $\gamma_{0}$, which marks the start of non-linearity in the response of the gel network to applied shear strains, is defined as the strain amplitude above which the elastic modulus $G^{\prime}$ decreases more than 5\% from its maximum value of $G^{\prime}_{0}$ in the amplitude sweep measurements. If $G^{\prime}_{0}$ is plotted \textit{vs.} increasing CaCl$_{2}$ concentration $C$, a power law increase ($G^{\prime}_{0} \sim C^{K}$) is observed (squares in Fig. 8). In the proposed egg-box model of pectin chain junctions \cite{braccini_biomacro}, Ca$^{2+}$ ions form cross-linking bridges between two adjacent non-methylated galacturonic units of two pectin chains. The 
number of cross-links, $N_{c}$, increases with increase in CaCl$_{2}$ concentration for a constant pectin concentration due to the increased availability of Ca$^{2+}$ ions. Following the argument of Lootens $\textit{et al.}$ \cite{looten_foodhydro}, if the elastic modulus $G^{\prime}$ depends only on the system entropy, $G^{\prime} \propto \nu$,where $\nu$ is the molar concentration of the elastically active network chains (EANC). For pectin gels with added calcium ions, $\nu = 2N_{c}- C_{p}/M_{n}$ \cite{clark_ijbm,clark_food,looten_foodhydro}, where $C_{p}$ and $M_{n}$ are, respectively, the concentration and the number average molar masses of the pectin chains. Increasing the CaCl$_{2}$ concentration, keeping the pectin concentration fixed, increases the value of $\nu$ due to an increase in $N_{c}$. This results in the observed increase of the elastic modulus $G^{\prime}$ of pectin gels with increasing CaCl$_{2}$ concentration. It is observed in Fig. 8 that while $G^{\prime}_{0}$ increases with added CaCl$_{2}$ concentrations, the non-linearity in $G^{\prime}$  starts to appear at lower strains in samples with higher added CaCl$_{2}$ concentrations. This is shown by vertical dotted lines in Fig. 7(b).\\
\begin{figure}
\begin{center}
\includegraphics[width=3.5in]{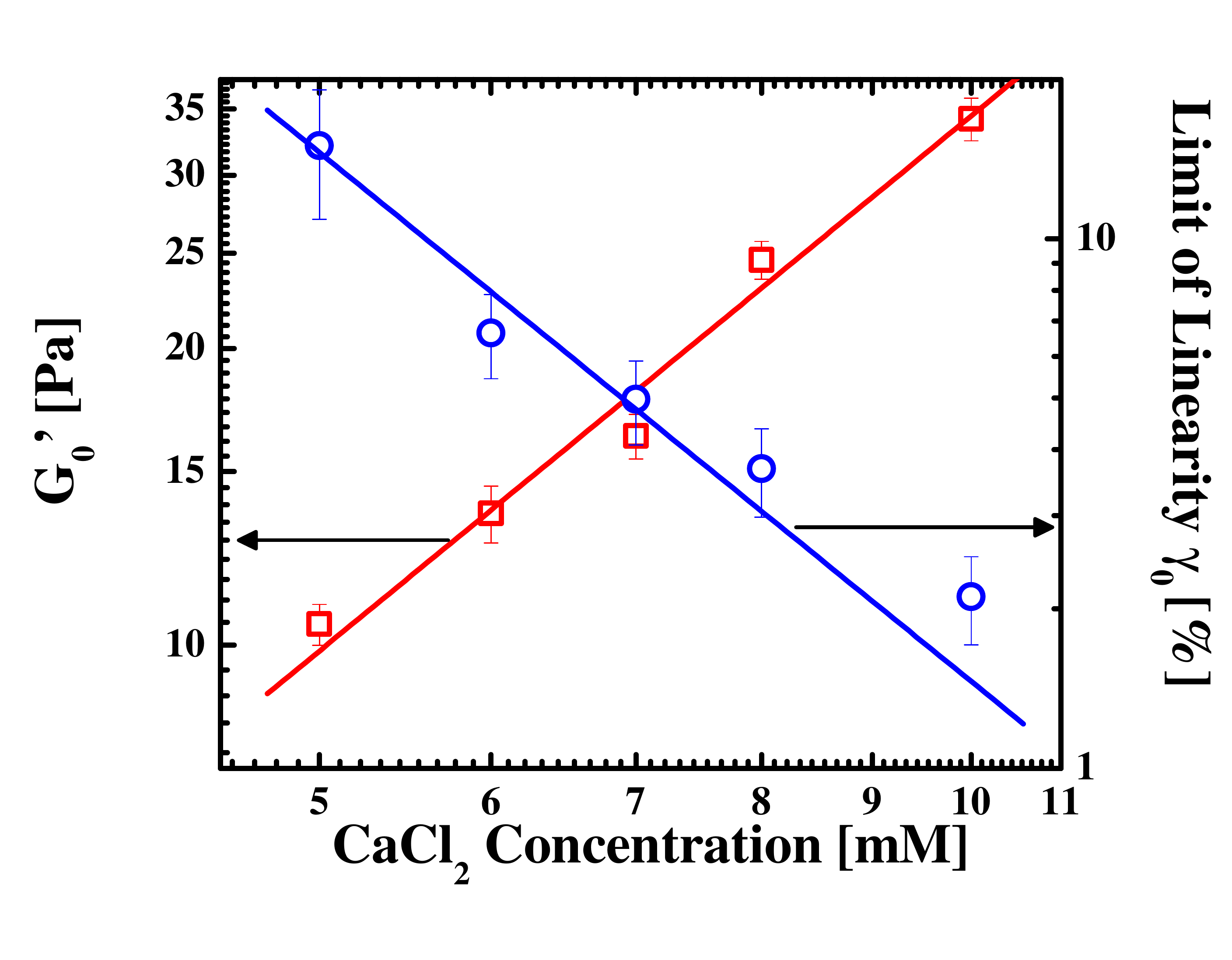}
\caption{ Plots of $G^{\prime}_{0}$ ($\square$)  and $\gamma_{0}$ ($\circ$) \textit{vs.} CaCl$_{2}$ concentration for 2.5 g/L pectin solutions with added salt. The corresponding power law fits ($G^{\prime}_{0} \sim C^{K}$ and $\gamma_{0} \sim C^{-L}$ with K= 1.78 and L=2.90) to the data are shown by solid lines.}
\label{FIG 8}
\end{center} 
\end{figure}
 \indent The limit of linearity, $\gamma_{0}$, and the estimated values of $G^{\prime}_{0}$ are plotted in Fig. 8 {\it vs.} CaCl$_{2}$ concentration.  A power law decrease in $\gamma_{0}$, ($\gamma_{0} \sim C^{-L}$, $\gamma_{\circ}$ represented by circles), is observed with increasing CaCl$_{2}$ concentration. The start of a non-linear elastic response in a cross-linked gel indicates the onset of plastic rearrangements in the network. As a lower critical stress is also observed with higher added CaCl$_{2}$ concentration (Fig. 7(a)), it can be concluded that the onset of non-linear behavior and the subsequent rupture of the gel network are accelerated significantly by an increase in CaCl$_{2}$ concentration. The yield strains, $\gamma_{ys}$, extracted from the peak positions of the plots of $G^{\prime\prime}$, are also observed to decrease with the added CaCl$_{2}$ concentration $C$. This data is plotted in the inset of Fig. 7(a). 
  \\  

 \indent The simultaneous lowering of $\sigma_{critical}$ and $\gamma_{\circ}$ for highly elastic pectin gels gives important information about the nature of the links that form between the pectin flocs. To explain the simultaneous power-law decrease of the limit of linearity and the power-law increase of the linear elastic modulus $G_{0}$ of elastic pectin gels with increasing salt concentration, a scaling approach proposed by Shih \textit{et al.} for fractal colloidal gels  in the strong link regime \cite{shih_prl} is followed here. It must be noted here that fractal concepts have previously been employed to analyse rheological data for biopolymer pectin gels \cite{bonn_science}, protein gel networks \cite{ikeda_lang, verhuel_lang} and colloidal gels \cite{shih_prl,wu_lang}. In lysozome protein gels, for example, the inverse strain-dependence of the upper limit of the linear viscoelastic region is intepreted to arise from the strong link behaviour of protein gels \cite{dasilva_jcis}. In all these previous studies, gelation was induced by changing the volume fractions of the gel-forming macromolecules. In this work, in contrast, gelation is triggered not by an increase in the volume fraction of pectin, but by an increased availability of Ca$^{2+}$ ions which enhances the formation of crosslinking bridges between pectin chains.

 It is observed in Fig. 3 that the mean relaxation time $<\tau_{R}>$ increases with increase in CaCl$_{2}$ concentration. This arises due to an increase in floc sizes before gelation. The increase in floc sizes with increasing salt concentration  is verified from cryo-SEM images (Fig. 4). For a reaction-limited fractal structure, simulation studies show that the radius of a cluster or floc, $\xi$, increases with cluster mass S according to a power law \cite{meakin_pra}. As the number of pectin chains increases inside a floc due to the increased formation of ion-induced crosslinks, it is reasonable to assume that the cluster mass S increases with salt concentration $C$. The floc size $\xi$ can therefore be expected to show the following power-law relation with CaCl$_{2}$ concentration
	 \[\xi\sim C^{M}  \tag{5}
 \]
  where $C$ is the concentration of CaCl$_{2}$ and M is a power law exponent. As floc sizes grow due to an increase in CaCl$_{2}$ concentration, the flocs start to behave like weak springs \cite{kantor_prl}, with the links within a floc becoming weaker due to a growth in floc sizes. For the pectin gels studied here, the links between the flocs (inter-floc links) can be characterized by higher elastic constants compared to the links within the individual flocs (intra-floc links). In this regime, when the applied stress is above a critical value, the links within the individual flocs can be assumed to break more easily than those between flocs. This is called `strong-link' regime \cite{shih_prl}. In this regime, the macroscopic elastic constant $G^{\prime}_{0}$ and the limit of linearity $\gamma_{0}$ depend on the individual floc elasticity $K_{\xi}$ and floc size $\xi$ according to following relations \cite{shih_prl}
	\[G^{\prime}_{0} \sim (K_{\xi}/\xi)  \tag{6}
	\]
	and 
	\[\gamma_{0} \sim (K_{\xi}\xi)^{-1} \tag{7}
	\]
		 The relevant calculation is shown in `Analysis and Calculations' section of Supporting Information. From Fig. 8, it is observed that $G^{\prime}_{0} \sim C^{K}$ and $\gamma_{0} \sim C^{-L}$. These power law relations are used along with Eqns. 6 and 7 to conclude that the individual floc elasticity $K_{\xi}$ also follows a power law with CaCl$_{2}$ concentration:
		
			\[K_{\xi} \sim C^{N}  \tag{8}
		\]
Here, $\frac{K+L}{2} = N$.  Increase in the individual floc elasticity $K_{\xi}$ with $C$ is clearly due to an increase in cross-linking density within a floc. As $\xi\sim C^{M}$ and $K_{\xi} \sim C^{N}$, Eqns. 6 and 7 can be written as $G^{\prime}_{0} \sim C^{N-M}$ and $\gamma_{0} \sim C^{-(N+M)}$. From the experimental data of Fig. 8 for 2.5 g/L pectin solutions with added salt, $N-M$  = 1.78 and $N+M$ = 2.90, which implies $M$ = 0.56 and $N$ = 2.34. This indicates that in a probable scenario wherein both $\xi$ and $K_{\xi}$ exhibit power-law variations with CaCl$_{2}$ concentration,  $\xi\sim C^{0.56}$ and $K_{\xi}\sim C^{2.34}$ for Ca$^{2+}$ induced 2.5 g/L pectin. The floc size $\xi$ is therefore weakly dependent on $C$. The individual floc elasticity $K_{\xi}$, in contrast, shows a much stronger dependence on $C$. These results are in agreement with the cryo-SEM images of 4(c) and 4(d), where the individual floc sizes do not appear to change significantly with $C$ when $C > C_{cr}$,  while the density of crosslinks shows a prominent increase. A stronger dependence of the elastic modulus on individual floc elasticity, rather than the floc size, was also observed  in the reaction limited aggregation regime of globular protein gels \cite{ikeda_lang}. \\
	\indent Finally, all the experiments reported earlier are repeated for aqueous pectin solutions of a higher concentration (5 g/L) to investigate the dependence of the power law exponents upon pectin concentration. From amplitude sweep experiments for different CaCl$_{2}$ concentrations $C$,  performed for a higher pectin concentration of 5 g/L, power-law behaviours of both the macroscopic elastic constant $G^{\prime}_{0}$ and the limit of linearity $\gamma_{0}$ are observed with increasing salt concentration. This data is shown in Supporting Information in Figs. S5 and S6. For 5 g/L pectin gels, $\xi$ and $K_{\xi}$ are described by the following relations: $\xi\sim C^{0.71}$ and $K_{\xi}\sim C^{2.52}$. Interestingly, even for the higher pectin concentration, it is observed that while $\xi$ is only weakly dependent on $C$, $K_{\xi}$ shows a much stronger dependence on salt concentration. For the 2.5 g/L pectin sample with 5 mM salt (squares in Fig. 7(b)),  $G^{\prime}_{\circ} \approx$ 7 Pa. Increasing both pectin and salt concentration by a factor of 2 (data for the 5 g/L pectin sample containing 10 mM salt, plotted in Fig. S5, is denoted by triangles)  increases the value of  $G^{\prime}_{\circ}$ is approximately  60 Pa. As the dependences of $\xi$ and $K_{\xi}$ on $C$ do not change considerably with pectin concentration, the large increase in  $G^{\prime}_{\circ}$ can be explained by considering a simultaneous increase in the number of pectin flocs and the density of inter-floc links when the pectin concentration is increased.

 \section{Conclusions}
Pectin is an important ingredient of plant cell walls, where ion-mediated pectin gels are known to contribute to the cell wall rigidity. Pectin is also often used in gel form  in the areas of food science and pharmaceuticals. These factors make a study of the mechanical properties of pectin gels extremely important. \\
\indent In this work, the gelation of pectin with increasing CaCl$_{2}$ concentrations and the rupture of gel networks due to the imposition of shear stresses have been studied systematically. Dynamic light scattering experiments are performed to study the formation of ion-induced pectin gel networks at different CaCl$_{2}$ concentrations. It is seen that the relaxation dynamics of the system shows a gradual slowdown with increase in CaCl$_{2}$ concentration due to the formation of larger pectin flocs. This increase is accompanied by the increase in complex viscosity, with the samples eventually entering a non-ergodic state, characterized by incomplete decays of the measured autocorrelation functions. The polydispersity of the flocs and the system inhomogeneity are seen to increase with CaCl$_{2}$ concentration before the onset of gelation. Cryo-SEM images are used to verify these results on gel formation qualitatively, while SLS studies confirm that the formation of pectin gels is governed by a 3-D  reaction-limited aggregation process. Next, systematic rheological studies are performed to understand the rupture of pectin gels. It is seen that when a constant stress is applied to pectin gels, the gel structures weaken due to the development of permanent strains and eventually break when the applied stress is above a critical value that decreases with salt (CaCl$_{2}$) concentration. Furthermore, the rheological studies reveal that increasing CaCl$_{2}$ concentration increases the elasticity of the gel, but decreases the limit of linearity of the strain response of the samples. This feature, together with the observation that the elastic moduli of the gels show a power-law dependence on salt concentration, demonstrates that pectin gels can be categorized to lie in a`strong link' regime in which inter-floc links are much stronger than intra-floc links. The existence of the strong link regime in pectin gels supports our observation that smaller critical stresses are required to break gel networks formed with larger CaCl$_{2}$ concentrations. In contrast to earlier reports where gelation was induced by increasing the volume fraction of the gel-forming molecule, our work investigates the process of ion-mediated gelation keeping the pectin concentration fixed. \\\\
\section{Additional Material}
\subsection{Supporting Information}
The following figures and calculations are supplied in a Supporting Information file. In part A, the  calculations of the scaling analyses of $G^{\prime}_{0}$ and $\gamma_{0}$ are described. Figure S1 shows the plots of the normalized autocorrelation functions $C(\tau)$ \textit{vs.} delay time $\tau$ for 2.5 g/L pectin solutions at different scattering angles. Figure S2 shows the plots of strain-evolutions during creep and recovery tests for 2.5 g/L pectin gels subjected to applied stress below $\sigma_{critical}$. Figure S3 shows the plots of strain-evolutions during creep and recovery tests for 2.5 g/L pectin gels for several applied stresses. This data is used to calculate $\sigma_{critical}$. Figure S4 shows the plots of loss modulus with strain amplitude which are used to measure the yield strains $\gamma_{ys}$ of  the 2.5 g/L pectin gels. Figure S5 shows the plots of $G^{\prime}$ {\it vs.} strain amplitude for different salt concentrations. This data is used to estimate the values of $G^{\prime}_{0}$ and $\gamma_{0}$ for 5 g/L pectin gels. Figure S6 shows the plots of $G^{\prime}_{0}$ and $\gamma_{0}$ \textit{vs.} CaCl$_{2}$ concentration for 5 g/L pectin gels .
\section{Acknowledgments}
The authors thank  D. Chelvan and A. Dhason for their assistance with the cryo-SEM measurement.
%\section{Conclusions}
%The conclusions section should come at the end of article. For the reference section, the style file rsc.bst can be used to generate the correct reference style.\footnote[4]{Footnotes should appear here. These might include comments relevant to but not central to the matter under discussion, limited experimental and spectral data, and crystallographic data.}
 %For footnotes in the main text of the article please number the footnotes to avoid duplicate symbols. e.g.  \footnote[num]{your text} the corresponding author \ast counts as footnote 1, ESI as footnote 2, e.g. if there is no ESI, please start at [num]=[2], if ESI is cited in the title please start at [num]=[3] etc. Please also cite the ESI within the main body of the text using \dag.


%\balance


\footnotesize{
%\bibliography{rsc} %your .bib file
%\bibliographystyle{rsc} %the RSC's .bst file
}

%-------------------------------------

%-------------------------------------
%\end{document}
%\newpage
%\begin{figure*}[!t]
%\begin{center}
%\includegraphics[width=10cm]{toc_hydrogel1.pdf}
%\caption{\bf Table of contents only}
%\end{center}
%\end{figure*}
%Flocs of pectin biomacromolecules, whose size and size polydispersities increase with added salt, connect to form a gel whose rheology is controlled mainly by the elasticity of the individual flocs.
\end{document}